\documentclass[12pt]{article}
\usepackage{graphicx}

\newcommand\pubnumber{Article 25 in eConf C1304143}
\newcommand\pubdate{\today}

\def\Stanford{Department of Physics\\
 Stanford University, Via Pueblo Mall, 94-301, Stanford, CA, USA}
\def\Jagellonian{Astronomical Observatory \\
 Jagellonian University, 31-044 Krakow, Poland}

\def\Title#1{\begin{center} {\Large #1 } \end{center}}
\def\Author#1{\begin{center}{ \sc #1} \end{center}}
\def\Address#1{\begin{center}{ \it #1} \end{center}}

\newcommand\pubblock{\rightline{\begin{tabular}{l} \pubnumber\\
         \pubdate  \end{tabular}}}
\newenvironment{Abstract}{\begin{quotation}  }{\end{quotation}}
\newenvironment{Presented}{\begin{quotation} \begin{center}
             PRESENTED AT\end{center}\bigskip
      \begin{center}\begin{large}}{\end{large}\end{center} \end{quotation}}
\def\Acknowledgements{\bigskip  \bigskip \begin{center} \begin{large}
             \bf ACKNOWLEDGEMENTS \end{large}\end{center}}

\def\beq{\begin{equation}}
\def\eeq#1{\label{#1}\end{equation}}
\def\eeqn{\end{equation}}

\def\beqa{\begin{eqnarray}}
\def\eeqa#1{\label{#1}\end{eqnarray}}
\def\eeqan{\end{eqnarray}}

\let\bar=\overbar

\def\Dslash{\not{\hbox{\kern-4pt $D$}}}
\def\dslash{\not{\hbox{\kern-2pt $\del$}}}

\def\msb{{\bar{\ssstyle M \kern -1pt S}}}

\begin{document}
\begin{titlepage}
\pubblock

\vfill
\Title{Study of luminosity and time evolution in X-ray afterglows of GRBs}
\vfill
\Author{Maria Giovanna Dainotti$^{1,2}$,Vahe' Petrosian$^{1}$,Jack Singal$^{1}$, Michal Ostrowski$^{2}$}
\Address{$^{1}$\Stanford\\ $^{2}$\Jagellonian }
\vfill
\begin{Abstract}
Gamma-ray bursts (GRBs), which have been observed up to redshifts $z \approx 9.5$ can be good probes of the early universe and have the potential of testing cosmological models. The analysis by Dainotti of GRB Swift afterglow lightcurves with known redshifts and definite X-ray plateau shows an anti-correlation between the \underline{rest frame} time when the plateau ends (the plateau end time) and the calculated luminosity at that time (or approximately an anti-correlation between plateau duration and luminosity). We present here an update of this correlation with a larger data sample of 101 GRBs with good lightcurves. 
Since some of this correlation could result from the redshift dependences of these intrinsic parameters, namely their cosmological evolution we use the Efron-Petrosian method to estimate the luminosity and time evolution and to correct for this effects to determine the intrinsic nature of this correlation. 
\end{Abstract}
\vfill
\begin{Presented}
GRB 2013 \\
the Seventh Huntsville Gamma-Ray Burst Symposium \\
Nashville, Tennessee, 14--18 April 2013
\end{Presented}
\vfill
\end{titlepage}
\def\thefootnote{\fnsymbol{footnote}}
\setcounter{footnote}{0}

\section{Introduction}

GRBs are the farthest sources, seen up to redshift $z=9.46$, and if emitting isotropically they are also the most powerful, (with $E_{iso} \leq 10^{54}$  erg  s$^{-1}$), objects in the Universe. In spite of the great diversity of their prompt emission lightcurves and their broad range spanning over 7 orders of magnitude of $E_{iso}$, some common features have been identified from investigation of their afterglow light curves.
A crucial breakthrough in this field has been the observation of GRBs by the \textit{Swift} satellite which provides a rapid follow-up of the afterglows in several wavelengths revealing a more complex behavior of the X-ray lightcurves than a broken power law generally observed before \cite{OB06,Sak07}. The {\it Swift} afterglow lightcurves manifest several segments. The second segment, when it is flat, is called the plateau emission.
A significant step forward in determining common features in the afterglow lightcurves was made by fitting them with an analytical expression \cite{W07}, called hereafter W07. 

This provides the opportunity to look for universal features that could provide a redshift independent measure of the distance. Dainotti et al. (2008, 2010), using the W07 phenomenological law for the lightcurves of long GRBs, discovered a formal anti-correlation  between the X-ray luminosity at the end of the plateau $L_X$ and the rest frame plateau end- time,  $T^{*}_a=T^{obs}_a/(1+z)$, (hereafter LT), described as :

\begin{equation}
\log L_X = \log a + b \log T^*_{a}, 
\label{feq}
\end{equation}
where $T^{*}_a$ is in seconds and $L_X$ is in erg/s. The normalization and the slope parameters $a$ and $b$ are constants obtained by the D'Agostini fitting method \cite{Dago05}.  
Dainotti et al. 2011a attempted to use the LT correlation as possible redshift estimator, but the paucity of the data and the scatter prevents from a definite conclusion at least for a sample of 62 GRBs. In addition, a further step to better understand the role of the plateau emission has been made with the discovery of new significant correlations between $L_X$, and the mean luminosities of the prompt emission, $<L_{\gamma,prompt}>$ \cite{Dainotti2011b}.
 
The LT anticorrelation is also a useful test for theoretical models such as the accretion models, \cite{Cannizzo09,Cannizzo11}, the magnetar models \cite{Dall'Osso,Bernardini2011,Bernardini2012,Rowlinson2010,Rowlinson2013}, the prior emission model \cite{Yamazaki09}, the unified GRB and AGN model \cite{Nemmen2012} and the fireshell model \cite{Izzo2012}. Furthermore, it has been recovered within also other observational correlations \cite{Ghisellini2008,Sultana2012,Qi2012}. Finally, it has been applied as a cosmological tool \cite{Cardone09,Cardone2010,Postnikov2013}.
Here, we study an updated sample of 101 GRBs and we investigate whether the LT correlation is intrinsic or induced by cosmological evolution of $L_X$ and $T^{*}_a$, and/or observational biases due to the instrumental threshold. This step is necessary to cast light on the nature of the plateau emission, to provide further constraints on the theoretical models, and possibly to assess the use of the LT correlation as a model discriminator.
In section \ref{Data} we describe the data and the results from correlation test carried using the raw data. In section \ref{intrinsic correlations} we use the EP method to determine the luminosity and time evolution corrections to finally determine the intrinsic correlation between $L_X$ and $T^*_{a}$.

\section{Lightcurve Data and raw correlations}\label{Data}

We have analyzed the sample of all GRB X-ray afterglows with known redshifts detected by {\it Swift} from January 2005 up to May 2011, for which the light curves include early X-ray data and therefore can be fitted by the W07 model. The source rest-frame luminosity in the {\it Swift} XRT bandpass, $(E_{min}, E_{max})=(0.3,10)$ keV at time $T_a$, is computed from the Equation:

\begin{equation}
L_X (E_{min},E_{max},T_a)= 4 \pi D_L^2(z) \, F_X (E_{min},E_{max},T_a) \times \textit{K},
\label{eq: lx}
\end{equation}

where $D_L(z)$ is the GRB luminosity distance \footnote{We assume a $\Lambda$CDM flat cosmological model with $\Omega_M = 0.291$ and $H_0 = 71  {\rm Km s}^{-1} {\rm Mpc}^{-1}$}, $F_X$ is the measured X-ray energy flux and $\textit{K}= (1+z)^{-1 +\beta_{a}}$ is the so called \textit{K}-correction for X-ray power law index $\beta_{a}$ \cite{Evans2009,Dainotti2010}.

\begin{figure}[htb]
\includegraphics[width=0.33\hsize,angle=0,clip]{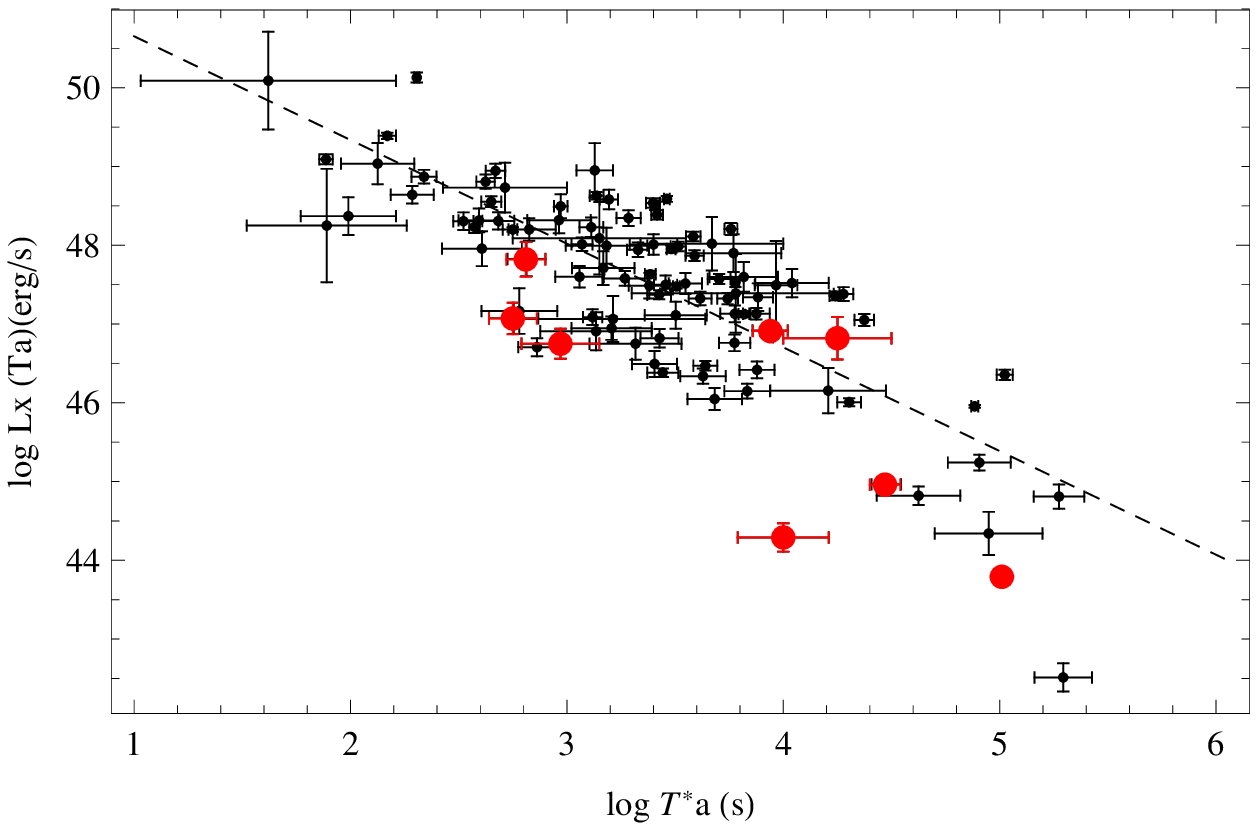}
\includegraphics[width=0.33\hsize,angle=0,clip]{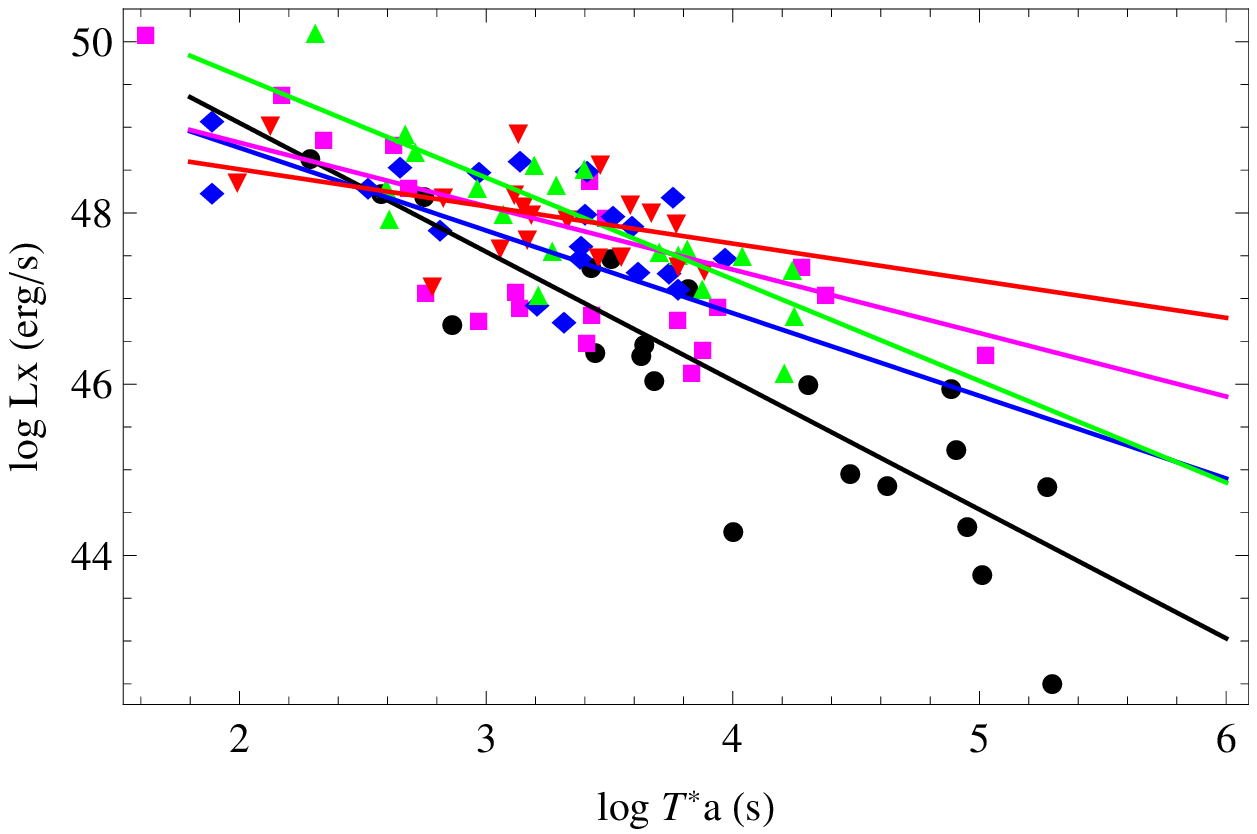}
\includegraphics[width=0.33\hsize,angle=0,clip]{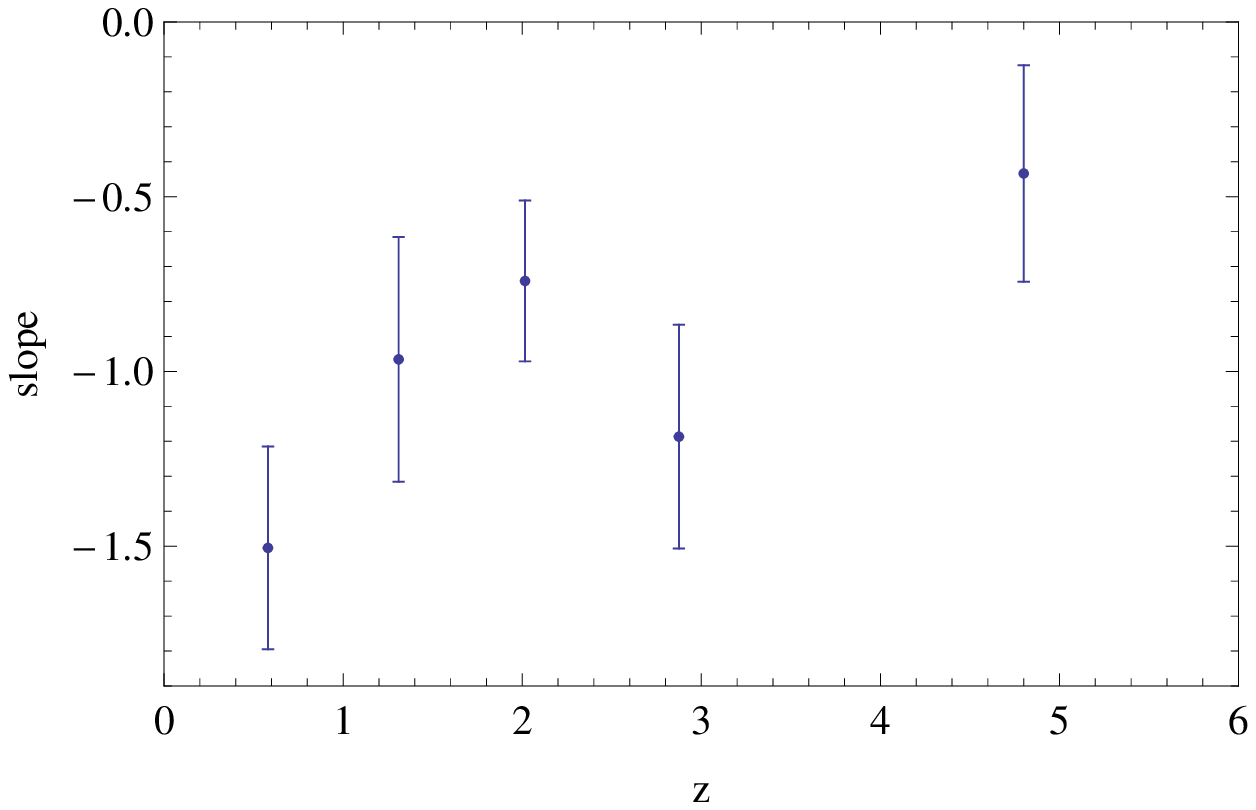}
\caption{{\bf Left Panel} $L_X$ vs $T^*_a$ distribution for the sample of 101 GRB afterglows with the fitted correlation shown by the dashed line. The red points are the IC bursts. {\bf Central Panel}: The same distribution divided in 5 equipopulated redshift bins shown by different colours: black for $z < 0.89$, magenta for $0.89 \leq z \leq 1.68$, blue for $1.68 < z \leq 2.45$, green $2.45 < z \leq 3.45$, red for $ z \geq 1.76$. Solid lines shows the fitted correlations. {\bf Right panel} The variation of the power law slope (and its error range) vith the mean value of the redshift bins.}
\label{fig1}
\end{figure}

Figure \ref{fig1}, left panel, shows the $L_X$-$T^*_a$ distribution of 101 GRBs with $0.08 \leq z \leq 9.4$ and includes afterglows of 93 long and 8 short bursts with extended emission \cite{nb2010}, called the Intermediate class (IC). For the whole sample without the IC we found the power law slope $b=-1.27 \pm _{-0.26}^{+0.18}$, while for the whole sample $b=-1.32 \pm _{-0.17}^{+0.18}$. The Spearman correlation coefficient for the larger sample ($\rho=-0.74$) is higher than $\rho=-0.68$ obtained for a subsample of 66 long duration GRBs analyzed in Dainotti et al. 2010. The probability of the correlation (of the 101 long GRBs) occurring by chance within an uncorrelated sample is $P \approx 10^{-18}$ . However, because both $L_X$ and $T^*_a$ depend on redshift ($L_X$ increasing and $T^*_a$ decresing with $z$) and the sample covers a broad redshift range all or part of the anticorrelation might be induced by these dependencies. It is therefore important to determine the extent of this effect and determine the true or intrinsic correlation.
In addition any cosmological evolution in $L_X$ and/or $T^*_a$ will affect the degree of the observed anti-correlation.
Fig.\ref{fig1}, central panel, shows the colour coded fitted lines. The distribution of the subsamples presents different power law slopes when we divide the whole sample into 5 redshift bins (see Dainotti et al. 2011 for a comparison with a smaller sample) thus having 20 GRBs in each subsample. The objects in different bins exhibit some separation into different regions of the $L_X$-$T^*_a$ plane. The results are shown in fig \ref{fig1} (central) with the fitted lines. In the right panel of Fig. \ref{fig1} we show the power slope of the redshift bins with the mean values of the redshif bins. 

\section{Determination of cosmological evolution and intrinsic correlations}\label{intrinsic correlations}

The first important step for determining the distribution of true correlations among
the variables is quantification of the biases introduced by the observational and sample selection effects.
In the case under study the selection effect or bias that distorts the statistical correlations are the flux limit and the temporal resolution of the instrument.
To account for these effects we apply the Efron \& Petrosian (1992) technique, already successfully applied for GRBs \cite{Petrosian2009}.

The EP method uses a modified version of the Kendall $\tau$ statistic to test the independence of variables in a truncated data.

\begin{figure}
\includegraphics[width=0.5\hsize,angle=0,clip]{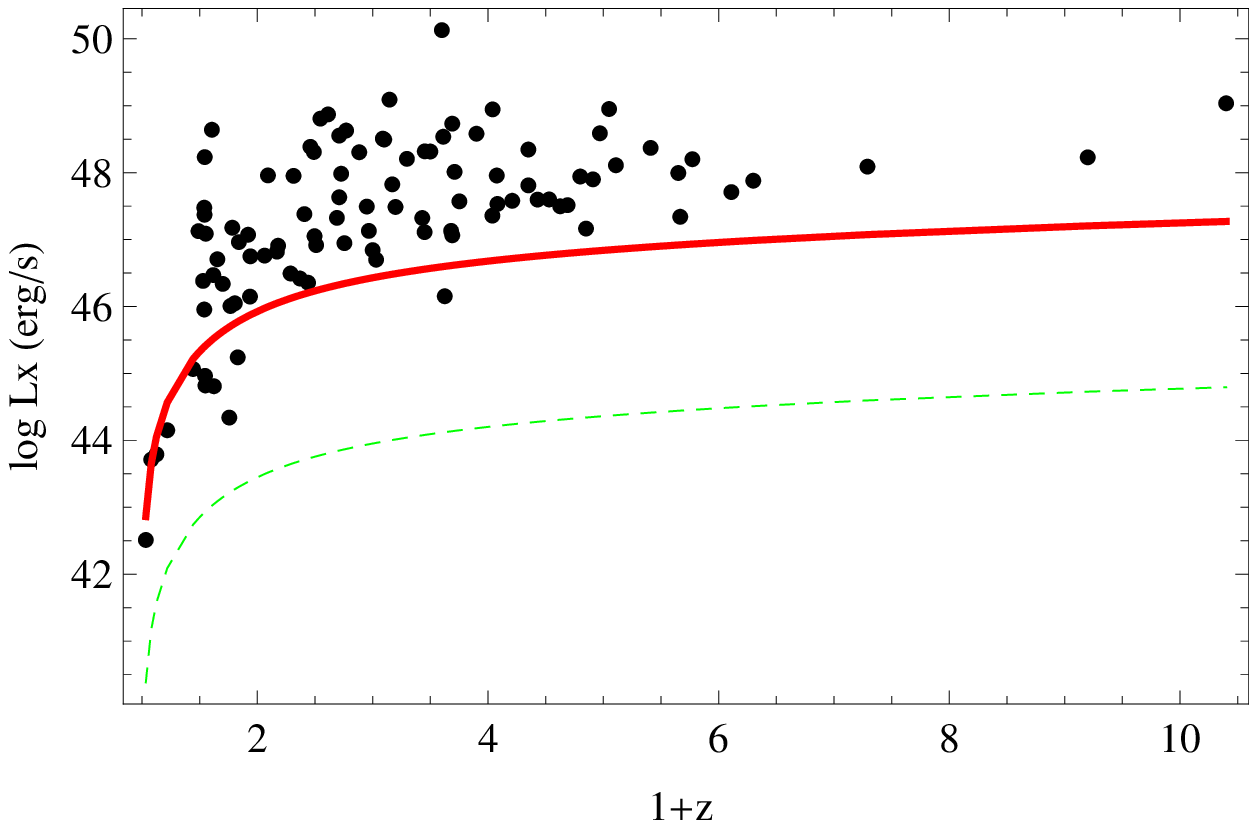}
\includegraphics[width=0.5\hsize,angle=0,clip]{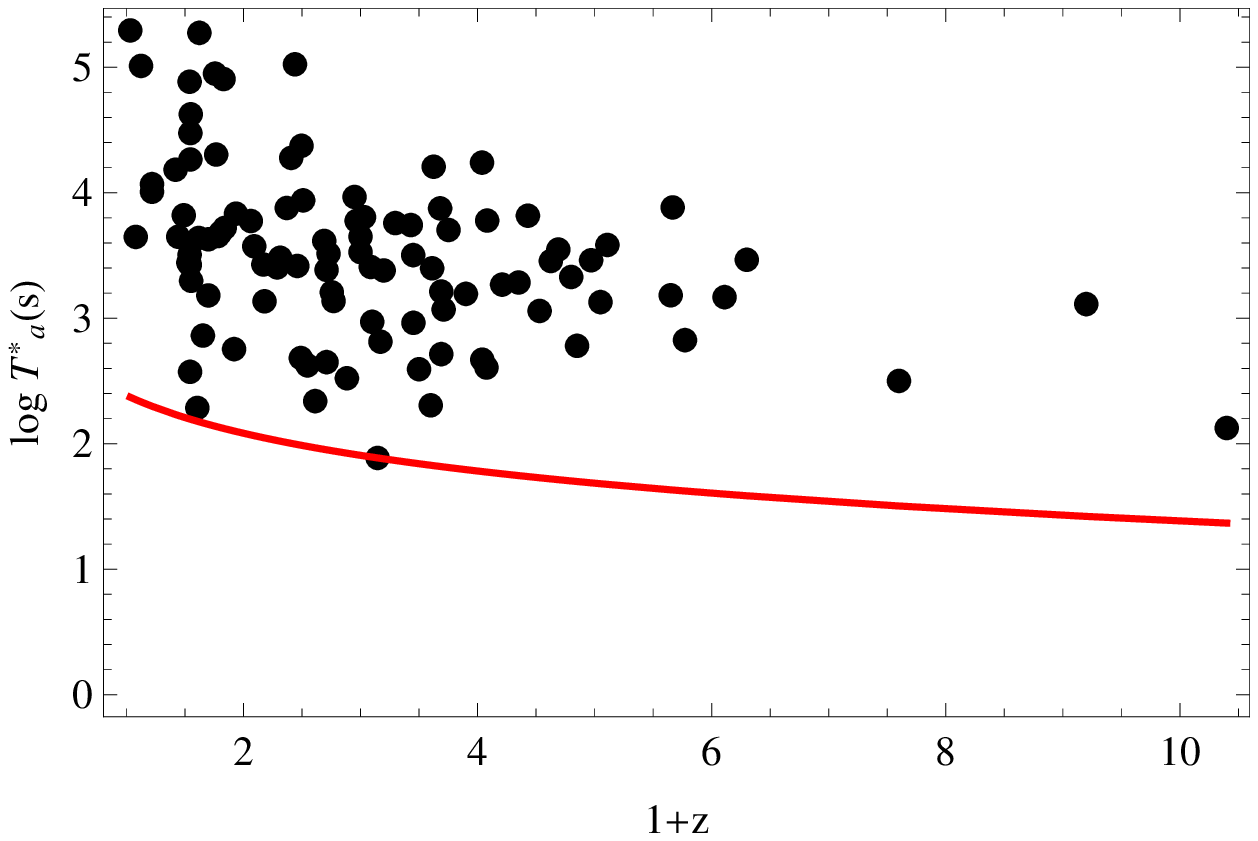}
\caption{{\bf Left Panel}: The bivariate distribution of $L_X$ and redshift with two different flux limits. The instrumental XRT flux limit,  $1.0 \times 10^{-14}$ erg cm$^{-2}$ (dashed green line) is too low to be representative of the flux limit, $1.5 \times 10^{-12}$ erg cm$^{-2}$ (solid red line) better represents the limit of the sample. 
{\bf Right panel}: The bivariate distribution of the rest frame time $T^*_a$ and the redshift. The chosen limiting value of the observed end-time of the plateau in the sample, $T_{a,lim}= 242$ s. The red line is the limiting rest frame time, $T_{a,{\rm lim}}/(1+z)$.}
\label{fig2}
\end{figure}

With this statistic, we find the parametrization that best describes the luminosity and time evolution.
This means that we have to determine the limiting flux, $F_{lim}$, which gives the minimum observed luminosity for a given redshift, $L_x=  4 \pi D_L^2(z) \, F_X K$ as shown in Fig. \ref{fig2}. The nominal limiting sensitivity of XRT, $F_{lim}=10^{-14}$ {\rm erg cm}$^{-2}$ ${\rm s}^{-1}$, is too low to describe the truncation of our sample, dashed line. This is because there is a limit in the plateau end times, $T^{*}_{a,{\rm lim}}= 242/(1+z)$ s, right panel of Fig. \ref{fig2}. Therefore, as pointed out by Cannizzo et al. 2011 this restriction increases the flux threshold to $10^{-12}$ erg cm$^{-2}$. Therefore, taking into account the above minimum plateau end time we have investigated several limiting fluxes to determine a good representative value while keeping an adequate size of the sample. We have chosen the limiting flux $F_{lim} = 1.5 \times $10$^{-12}$ erg cm$^{-2}$, shown by the red solid line, which allows 90 GRBs in the sample.

\begin{figure}
\includegraphics[width=0.50\textwidth,height=0.48\textwidth]{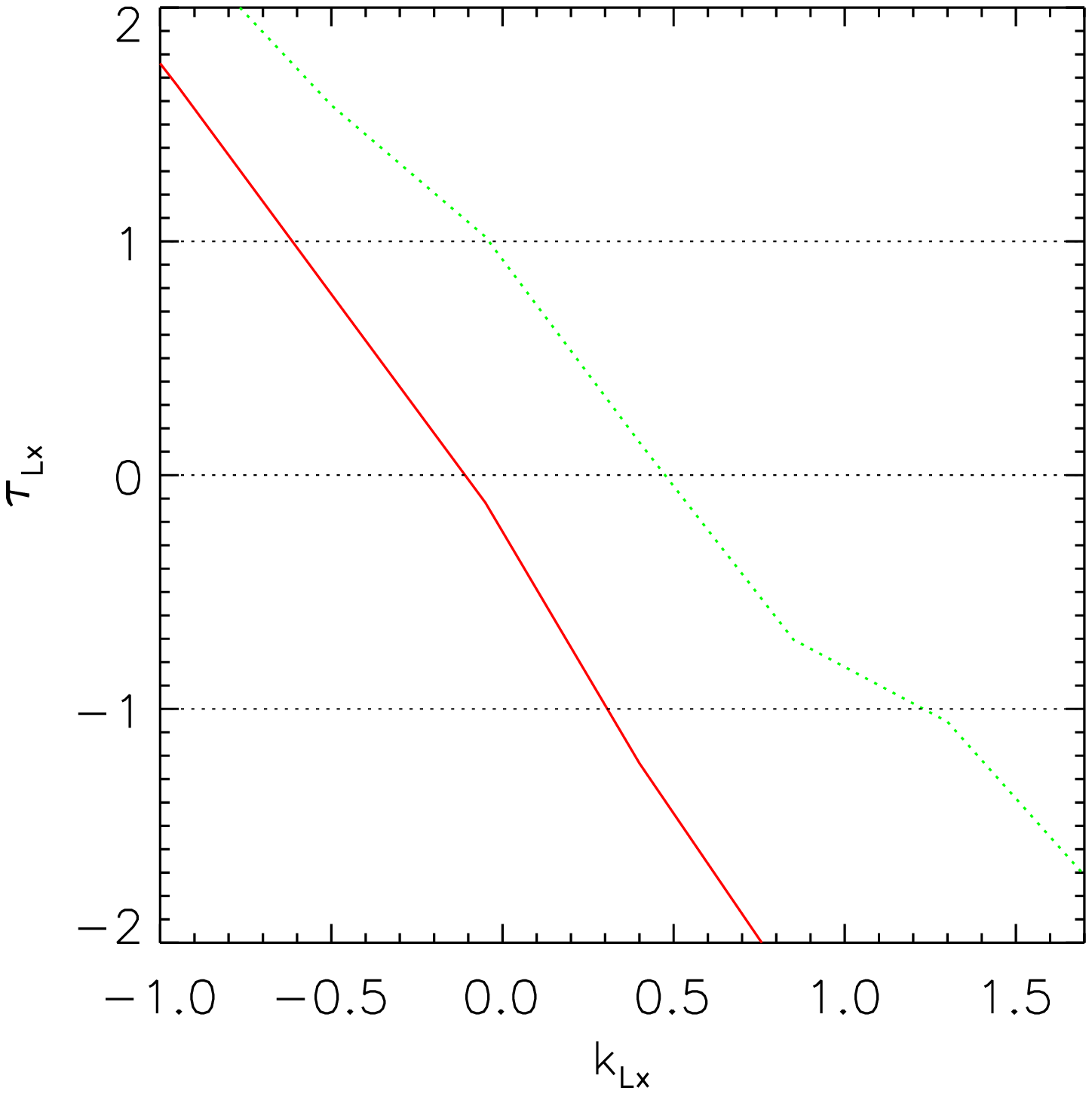}
\includegraphics[width=0.50\textwidth,height=0.50\textwidth]{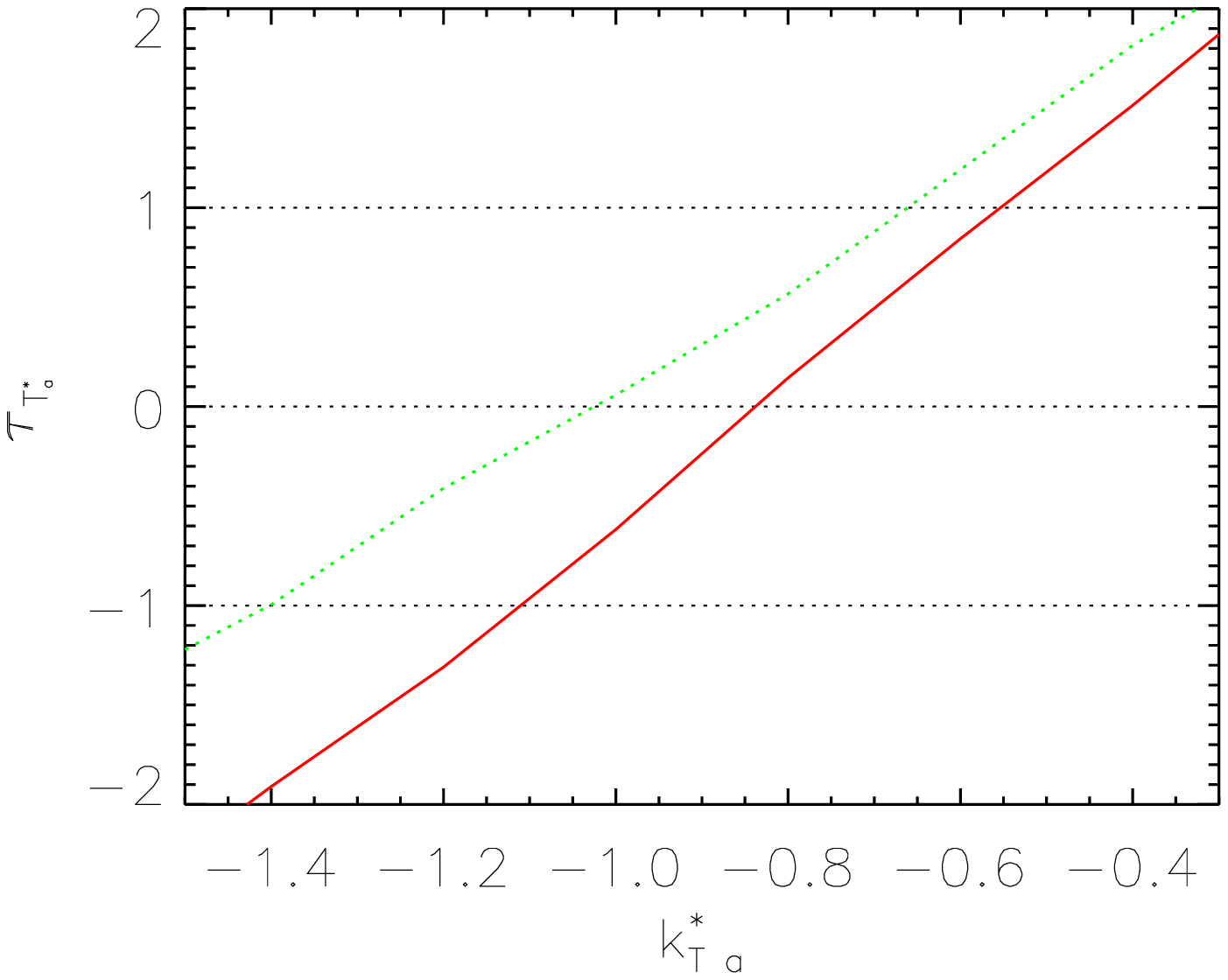}
\caption{ {\bf Left:} Test statistic $\tau$ vs. $k_{L_x}$, the luminosity evolution. Right panel Test statistic $\tau$ vs. $k_{T^{*}_a}$, the time evolution. The red line represents the full sample of 101 GRBs, while the green line represents the small sample of 47 GRBs in common with the previous sample of 77 GRBs.}
 \label{Fig.3}
\end{figure}

The first step required for this kind of investigation is the determination of whether
the variables $L_X$ and $T^*_a$, are correlated with redshift or are
statistically independent of it. For example, the correlation between $L_X$ and the redshift, $z$, is what we
call luminosity evolution, and independence of these variables
would imply absence of such evolution. 
The EP method prescribed how to remove the correlation by defining
new and independent variables.

We determine the correlation functions, g(z) and f(z) when determining the evolution of $L_X$ and $T^{*}_a$  so that de-evolved variables $L'_{X} \equiv L_X/g(z)$ and $T'_a \equiv T^*_a/f(z)$ are not correlated with z. 
The evolutionary function are parametrized by simple correlation functions

\begin{equation}
g(z)=(1+z)^{k_{Lx}}, f(z)=(1+z)^{k_{T^{*}a}}
\label{lxev}
\end{equation}

With the specialized version of Kendell's $\tau$ statistic, the values of $k_{L_x}$ and $k_{T^{*}a}$ for which $\tau_{L_x} = 0$ and $\tau_{T^{*}a} = 0$ are the ones that best fit the luminosity and plateau end time evolution respectively, with the 1$\sigma$ range of uncertainty given by $| \tau_x | \leq 1$. Plots of $\tau_{L_x}$ and $\tau_{T^{*}a}$ versus $k_{L_x}$ and $\tau_{T^{*}a}$ are shown in Fig. 3. With $k_{L_x}$ and $k_{T^{*}a}$ we are able to determine the de-evolved observables $T{'}_a$ and $L{'}_X$. 

We evident there is no discernable luminosity evolution, $k_{L_x}=-0.05_{-0.55}^{+0.35}$, but there is a significant evolution in $T^{*}_a$, $k_{T^{*}a}=-0.85_{-0.30}^{+0.30}$.
 
The further application of the EP method in the new parameter space of the $L'_{X}$ and $T^{'}_a$ variables enable us to determine the intrinsic slope of the LT correlation, $1.07$ and to establish that the correlation is significant at 12 $\sigma$ level. With the EP method we are able both to overcome the problem of selection effects and to determine the intrinsic value of the slope, because we removed the induced correlation by observables due to the time evolution and luminosity evolution dividing the respective time and luminosity for the respective evolution functions.
Therefore, the presented analysis, with the intrinsic value of the power law slope of the LT correlation, provides new constraints for physical models of GRB explosion mechanisms. 
With this new determination of the correlation power law slope we can discuss the consequences of these findings for GRB physical models.

\Acknowledgements
This work made use of data supplied by the UK Swift Science Data Centre at the University of Leicester. I am grateful to Richard Willingale and Paul O'Brien for fruitful discussions.

\end{document}